\documentclass[sigconf]{acmart}

\AtBeginDocument{%
  }

\setcopyright{acmlicensed}

\copyrightyear{2024}
\acmYear{2024}
\setcopyright{rightsretained}
\acmConference[CHI EA '24]{Extended Abstracts of the CHI Conference on Human Factors in Computing Systems}{May 11--16, 2024}{Honolulu, HI, USA}
\acmBooktitle{Extended Abstracts of the CHI Conference on Human Factors in Computing Systems (CHI EA '24), May 11--16, 2024, Honolulu, HI, USA}
\acmDOI{10.1145/3613905.3650733}
\acmISBN{979-8-4007-0331-7/24/05}

\usepackage{hyperref}
\usepackage{enumitem}
\usepackage{subcaption}
\usepackage{makecell}
\DeclareCaptionLabelFormat{subfigure}{(#2)}
\begin{document}

\title[Hapster: Apple Watch Haptics for Student Feedback]{Hapster: Using Apple Watch Haptics to Enable Live Low-Friction Student Feedback in the Physical Classroom}

\author{Oleg Golev}
\authornote{All four authors contributed equally to this research.}
\affiliation{%
  \institution{Princeton University}
  \city{Princeton}
  \state{New Jersey}
  \country{USA}
  \postcode{08540}
}
\email{ogolev@princeton.edu}

\author{Michelle Huang}
\authornotemark[1]
\affiliation{%
  \institution{Princeton University}
  \city{Princeton}
  \state{New Jersey}
  \country{USA}
  \postcode{08540}
}
\email{mh52@princeton.edu}

\author{Kritin Vongthongsri}
\authornotemark[1]
\affiliation{%
  \institution{Princeton University}
  \city{Princeton}
  \state{New Jersey}
  \country{USA}
  \postcode{08540}
}
\email{kritinv@princeton.edu}

\author{Chanketya Nop}
\authornotemark[1]
\affiliation{%
  \institution{Princeton University}
  \city{Princeton}  
  \state{New Jersey}
  \country{USA}
  \postcode{08540}
}
\email{cnop@princeton.edu}

\author{Andrés Monroy-Hernández}
\affiliation{%
  \institution{Princeton University}
  \city{Princeton}
  \state{New Jersey}
  \country{USA}
  \postcode{08540}
}
\email{andresmh@princeton.edu}

\author{Parastoo Abtahi}
\affiliation{%
  \institution{Princeton University}
  \city{Princeton}
  \state{New Jersey}
  \country{USA}
  \postcode{08540}
}
\email{parastoo@princeton.edu}


\begin{abstract}
    The benefits of student response systems (SRSs) for in-person lectures are well-researched. However, all current SRSs only rely on a visual interface to relay information to the instructor. We describe the design and evaluation of Hapster, a prototype system that uses an Apple Watch to deliver live, aggregated student feedback to the instructor via both visual and vibro-tactile modalities. We evaluated this system with 6 instructors and 155 students at a U.S. university. Participants reported that the system was effective at delivering live student feedback and facilitating better engagement from both the instructor and the students. However, instructors also noted several challenges with differentiating and perceiving the haptic sequences while lecturing. We conclude by discussing the tradeoff between system flexibility and abuse potential while identifying opportunities for further research regarding accessibility, content moderation, and additional interaction modalities. Our results suggest that haptics can be used as an effective live feedback mechanism for instructors in the physical classroom.
    
\end{abstract}




\begin{teaserfigure}
  \includegraphics[width=\textwidth]{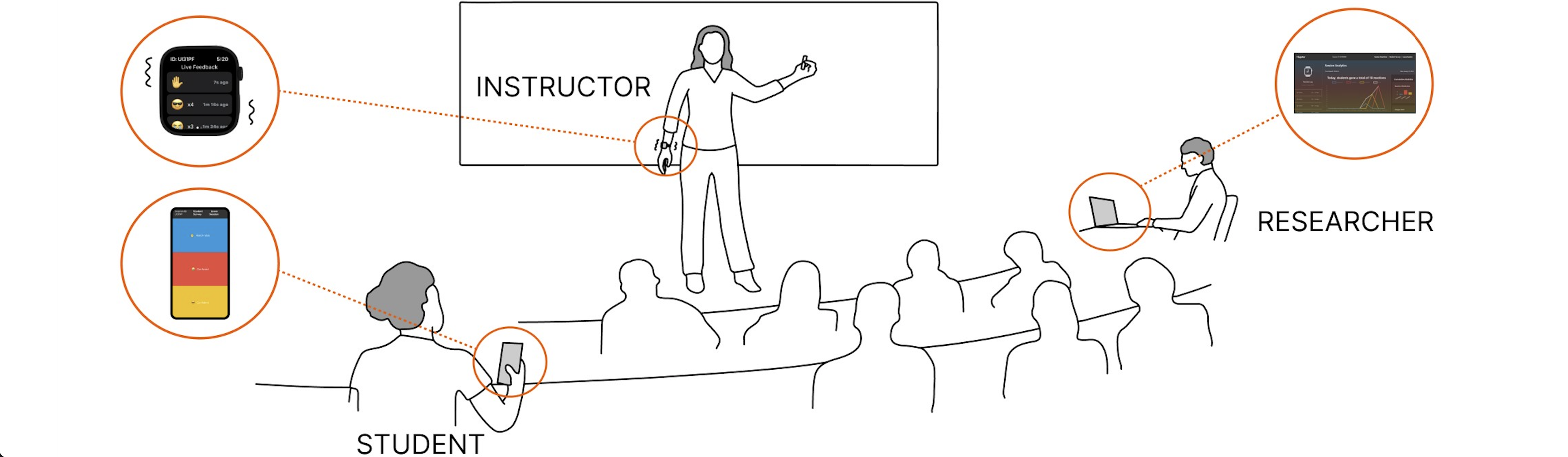}
  \Description[A classroom with an instructor, a researcher, and several students using Hapster.]{
    Hapster used in a classroom. An instructor lectures to a class, wearing an Apple Watch receiving notifications. A student uses the web interface on a mobile device. A researcher observes the Hapster dashboard on a laptop.}
  \caption{Hapster used in the classroom: students submit reactions; instructor receives visual and haptic live feedback that corresponds to the anonymous and aggregated student reactions; researcher observes the analytics dashboard.\vspace{0.3cm}}
  \label{fig:teaser}
\end{teaserfigure}


\maketitle

\section{Introduction}

Academic instructors often use technology to help adapt their teaching to student needs and improve student engagement. Student response systems (SRSs) such as Slido\footnote{slido.com}, Mentimeter\footnote{mentimeter.com}, and Wooclap\footnote{wooclap.com}, have been developed for those reasons. However, these SRSs are limited to the visual modality and thus demand active visual attention from instructors. This is particularly problematic if an instructor's teaching method, such as writing on a blackboard, requires them to face away from the students or the SRS visual interface for extended periods of time. Furthermore, these existing systems do not notify the instructor of new student input, introducing delay between when a student uses the SRS and when the instructor acknowledges that student input.

To address these issues, we developed Hapster, a novel student response system for the Apple Watch that not only provides instructors with a visual interface to view student reactions but also plays a haptic vibration sequence uniquely associated with each supported student reaction as soon as that reaction is sent. Multiple reactions of the same type that are sent within a 10-second timeframe are aggregated, with the count displayed visually and the haptic sequence playing only once to reduce noise. Hapster consists of two components: a student-facing web application through which students can provide anonymous reactions to their instructor and a watchOS application installed on the instructor's Apple Watch. Hapster currently supports \textit{Confused}, \textit{Hand-Raise}, and \textit{Confident} reactions, because these three reactions were most requested by instructors in our preliminary survey (see Section \hyperref[subsec:rxns]{4.1} for details).

We evaluated Hapster in six courses during regularly-scheduled lecture time. Students reported that the system was more effective than others (like Slido and Mentimeter) in allowing them to communicate the three reactions to their instructor. The system's anonymous nature also encouraged students to participate in cases where they usually would not feel comfortable doing so without an SRS. We found that the Hapster system was generally easy to use, facilitated higher student engagement and better instructor engagement, and gave instructors valuable information to better regulate their lecture pace. We also synthesized multiple improvements to implement for the next iteration of the system, mostly to reduce the system's cognitive load for instructors as well as to prevent student abuse of the system.

\section{Related Work}

\subsection{Student Response Systems}
Researchers have found that instructors improve their teaching quality through constructive feedback \cite{StudentFeedbackToImproveTeaching}. Specifically, immediate feedback is generally more effective than delayed feedback in facilitating learning \cite{ImmediateVersusDelayedFeedback}. To that purpose, developers created various SRSs to collect and display live student feedback for instructors. SRSs can not only help instructors gauge their student audience, but they have also proven to improve student motivation, activity levels, and engagement, with ease of use and anonymity cited as the most common student-reported reasons for these improvements since ease of use lowers the barrier to interaction, and anonymity removes the fear of negative peer evaluation and the associated anxiety \cite{CallOnMe}. Studies have reaffirmed these trends for popular applications like Slido \cite{Slido1, Slido2, Slido3} and Mentimeter \cite{Mentimeter3, Mentimeter2, Mentimeter1}, as well as systems developed to target specific aspects of speaker-audience interaction, such as the Live Interest Meter (LIM) App for aggregating and visualizing feedback from a big audience \cite{LIMApp} or the Mudslide system for feedback in a flipped classroom model \cite{Mudslide}. However, while these systems successfully increase student engagement, they all rely on sustained or frequent visual attention from the instructor. Since this is not always possible, the instructors' reactions to student feedback can be significantly delayed, especially if the instructors primarily use the blackboard and thus must face away from the student audience or any live visual feedback via an SRS. 

No previous work in SRSs has addressed alternatives to visual interfaces as modalities of feedback, but there is clear potential that haptic technologies can effectively reduce communication latencies in this space \cite{TactileTemporalities}. Thus we build and evaluate an SRS that does not solely rely on a visual interface by also providing a vibrotactile (haptic) modality.

\subsection{Haptic Feedback}

Studies in other domains have shown the value of haptic feedback in conveying information \cite{DiscriminationMethods}, as people can learn and distinguish between up to 10 distinct haptic sequences via smartwatches \cite{ActiVibe}, phones \cite{AssistBlind}, and other wearable devices, such as a vest with actuators placed on the back \cite{ArmAndBack}. Haptics have also been evaluated for vastly different audiences and applications. For example, prior research exploring the potential of phone-mediated haptic feedback for aiding people with visual impairments in indoor navigation found that participants identified the majority of the patterns in less than 4 seconds, with four patterns achieving over 80\% recognition and two patterns surpassing 90\% \cite{AssistBlind}. Other research found that well-designed haptic patterns, as played through a multi-actuator wrist device, with even a short duration of 0.3 s and 0.5 s can achieve recognition accuracies of around 80\% and 90\%, respectively \cite{xu2024haptics}.

Therefore, we see that well-designed haptic sequences can be quickly and accurately identified to convey information. Despite such promising findings, there is a noticeable absence of research on the applications of haptic feedback in classroom settings. Based on previous research, we expect haptics to perform well in a classroom scenario, where students would be able to use haptic patterns to deliver live discrete information to instructors. Thus, we designed Hapster to explore this gap.

\section{Hapster}

\begin{figure}[!htb]
  \centering
  \begin{minipage}{0.9\linewidth}
    \centering
    \begin{subfigure}[b]{0.5\textwidth}
     \centering
      \includegraphics[width=.9\linewidth]{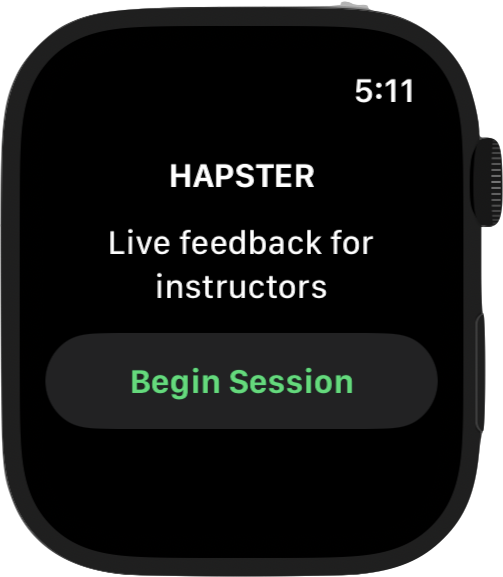}
      \Description[App home screen]{Hapster title page, with a 'Begin Session' button}
      \caption{}
    \end{subfigure}%
     \begin{subfigure}[b]{0.5\textwidth}
       \centering
      \includegraphics[width=.9\linewidth]{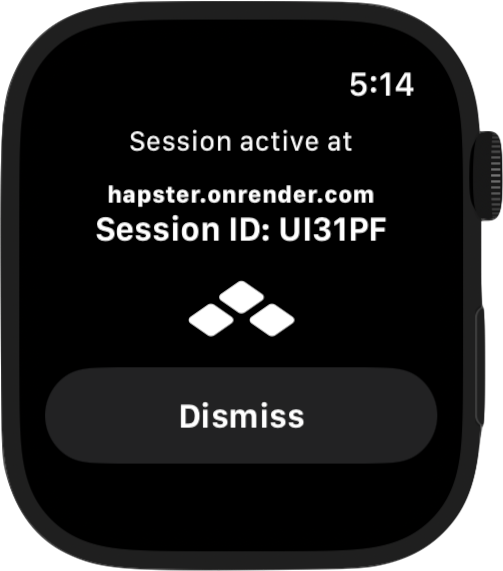}
      \Description[Session ID screen]{Active session screen, displaying the generated Session ID number, with 'Dismiss' button at the bottom}
      \caption{}
    \end{subfigure}
     \begin{subfigure}[b]{0.5\textwidth}
       \centering
      \includegraphics[width=.9\linewidth]{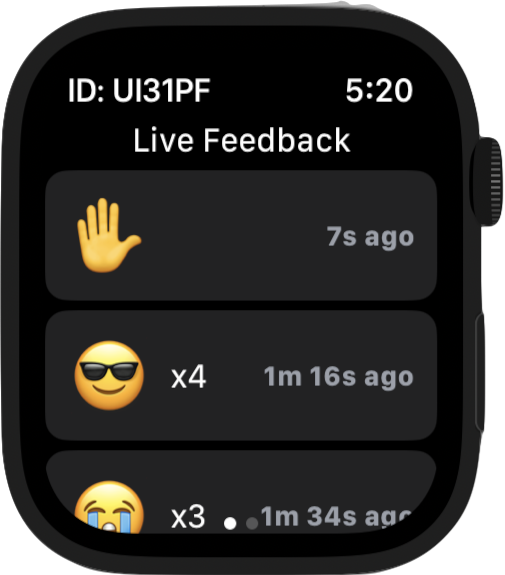}
      \Description[Live feedback log]{Log with a hand-raise emoji, sunglasses emoji, and crying emoji.}
      \caption{}
    \end{subfigure}%
     \begin{subfigure}[b]{0.5\textwidth}
       \centering
      \includegraphics[width=.9\linewidth]{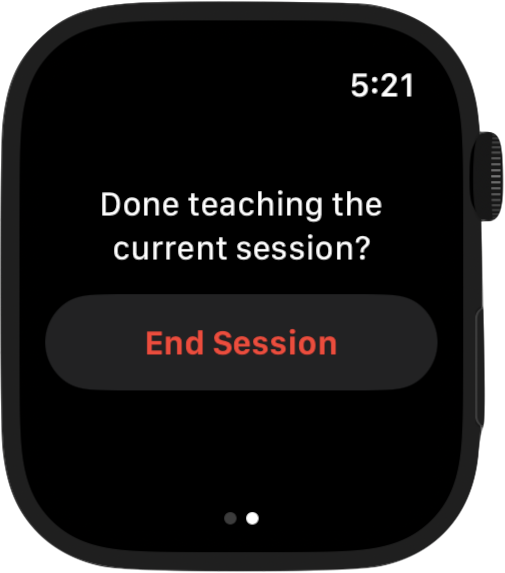}
      \Description[End session screen]{'End Session' button}
      \caption{}
    \end{subfigure}
  \end{minipage}%

  \Description[Hapster Apple Watch interfaces]{}
  \caption{Instructor-facing watchOS interface. (a) The instructor taps the \textit{Begin Session} button on the home page. (b) The instructor views a confirmation message displaying the new session ID, which is then dismissed. (c) The instructor views the live reaction log, which contains emojis indicating the reactions that the students sent. (d) The instructor taps \textit{End Session} to end the session.}
  \label{watchosapp}
\end{figure}

The Hapster system consists of two core components: a watchOS application installed on the instructor's Apple Watch (Figure \ref{watchosapp}) and a student-facing web application (Figure \ref{webapp}). We also designed a researcher-facing web interface (Figure \ref{researcher-dashboard}) for live session monitoring during system studies. We created a REST API service to support all these components, and we chose Firebase Firestore to store session data (all reactions with timestamps and anonymous user IDs) upon session termination.

We developed the web components using React and deployed them on Render as a static site. We built the watchOS application using SwiftUI and deployed it using TestFlight. Finally, we wrote the API service using FastAPI and also deployed it to Render. 

\subsection{Instructor-Facing watchOS Application}

\begin{figure}[!htb]
  \centering
  \begin{minipage}{0.9\linewidth}
    \centering
    \begin{subfigure}[b]{0.5\textwidth}
     \centering
      \includegraphics[width=.9\linewidth]{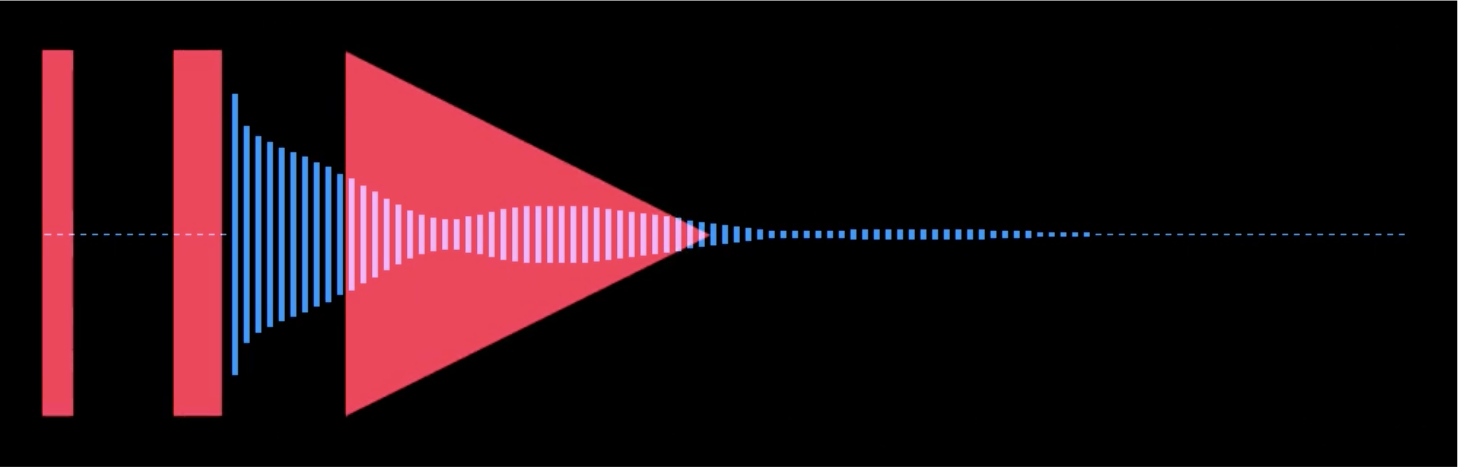}
      \Description[Hand-raise haptic waveform]{Haptic waveform with 2 rectangular and 1 triangular red overlay}
      \subcaption{\textit{Hand-raise}}
        \label{notif}
    \end{subfigure}%
     \begin{subfigure}[b]{0.5\textwidth}
       \centering
      \includegraphics[width=.9\linewidth]{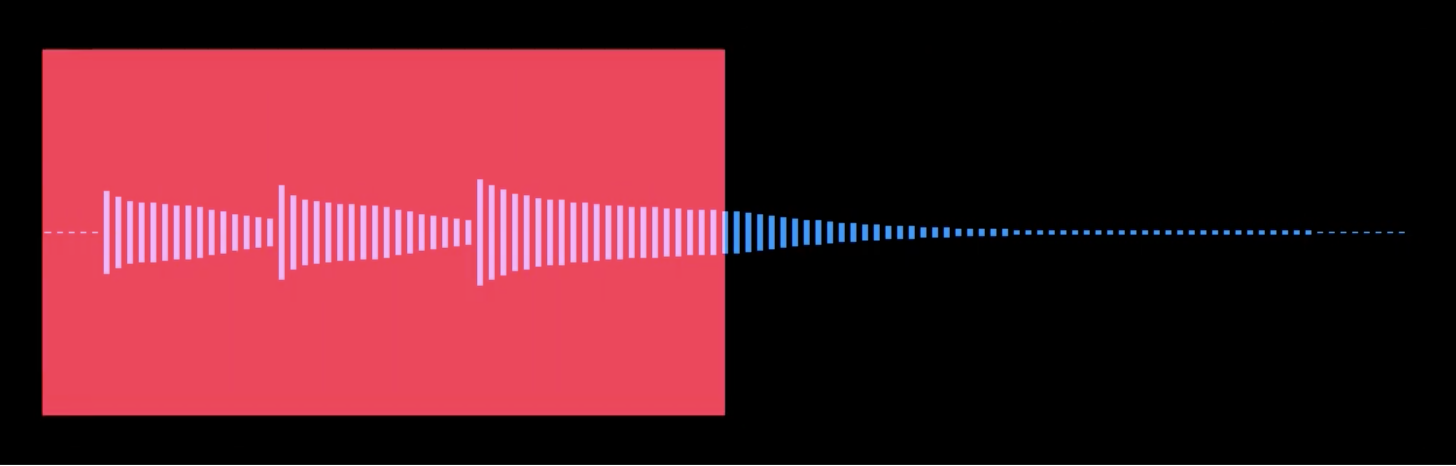}
       \Description[Confused haptic waveform]{Haptic waveform with 1 rectangular red overlay spanning half the waveform}
        \subcaption{\textit{Confused}}
        \label{fail}
    \end{subfigure}
     \begin{subfigure}[b]{0.5\textwidth}
       \centering
      \includegraphics[width=.9\linewidth]{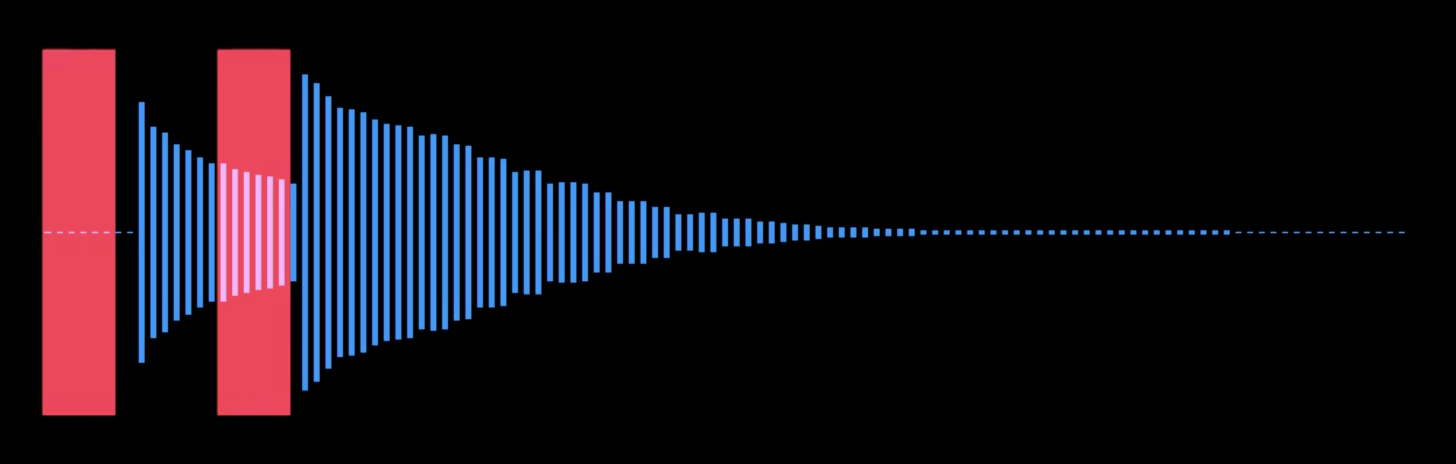}
       \Description[Confident haptic waveform]{Haptic waveform with 2 rectangular red overlays}
      \subcaption{\textit{Confident}}
        \label{up}
    \end{subfigure}%
  \end{minipage}%
  \Description[Three haptic pattern waveforms]{Haptic waveforms with vibration overlays for the hand-raise, confused, and confident reactions}
  \caption{Each reaction is assigned a predefined watchOS haptic pattern \cite{apple} played four times in quick succession to form a haptic sequence. The blue waveform corresponds to the sounds emitted, and the red overlay indicates the produced vibration. (a) \textit{WKHapticType.notification} for \textit{Hand-Raise} feels like a heartbeat. (b) \textit{WKHapticType.failure} for \textit{Confused} feels like a strong vibration. (c) \textit{WKHapticType.directionUp} for \textit{Confident} feels like two short ticks.}
  \label{waveforms}
\end{figure}

To use the system, the instructor first opens the Hapster application on their Apple Watch and creates a new session, generating a random session ID. The ID is then given to the students who enter it on their devices into the web application. The instructor sees all reactions in the order they arrive on a visual log, with each type of reaction represented as a unique emoji and accompanied by a unique haptic sequence (waveforms shown in Figure \ref{waveforms}). Multiple reactions of the same type sent within a 10-second time frame are grouped together and displayed alongside an aggregated count, with the corresponding haptic sequence playing only once.

We opted to develop the instructor-facing side of the application on the Apple Watch due to its widespread adoption and its vibrotactile capabilities. Likewise, its predefined set of haptic patterns was sufficient for our exploratory purposes. For instance, the lack of customizable vibration frequency in watchOS was not an obstacle since people are not significantly sensitive to frequency \cite{ArmLocalization}. On the other hand, strong vibrations and skin-squeezing sensations help maximize differentiation \cite{TactileMasking}, with bigger amplitude and longer duration being most effective at enabling faster pattern recognition \cite{EffectivenessOfHapticsUnderCognitiveLoad}. Thus, we set \textit{Confused} and \textit{Hand-Raise} to use haptic patterns with strong vibrations as they are more intrusive. For the \textit{Confident} reaction, we assigned a pattern of short low-intensity vibrations, meant to be less intrusive.

\subsection{Student-Facing Web Application}

Students may access the web interface using any browser-compatible device. The students first enter a session ID provided to them by their instructor before being redirected to the main interface with the three reaction options. When tapping on any of the reactions, the student’s selection is confirmed by showing them a flying emoji animation of the corresponding reaction, and their reaction is transmitted directly to the instructor's Apple Watch. After tapping on a reaction, a 20-second cooldown begins, disabling that button for that duration to prevent spamming.

\setcounter{figure}{4}
\begin{figure*}[!ht]
    \centering
    \begin{minipage}{.22\textwidth}
        \centering
        \includegraphics[scale=0.25]{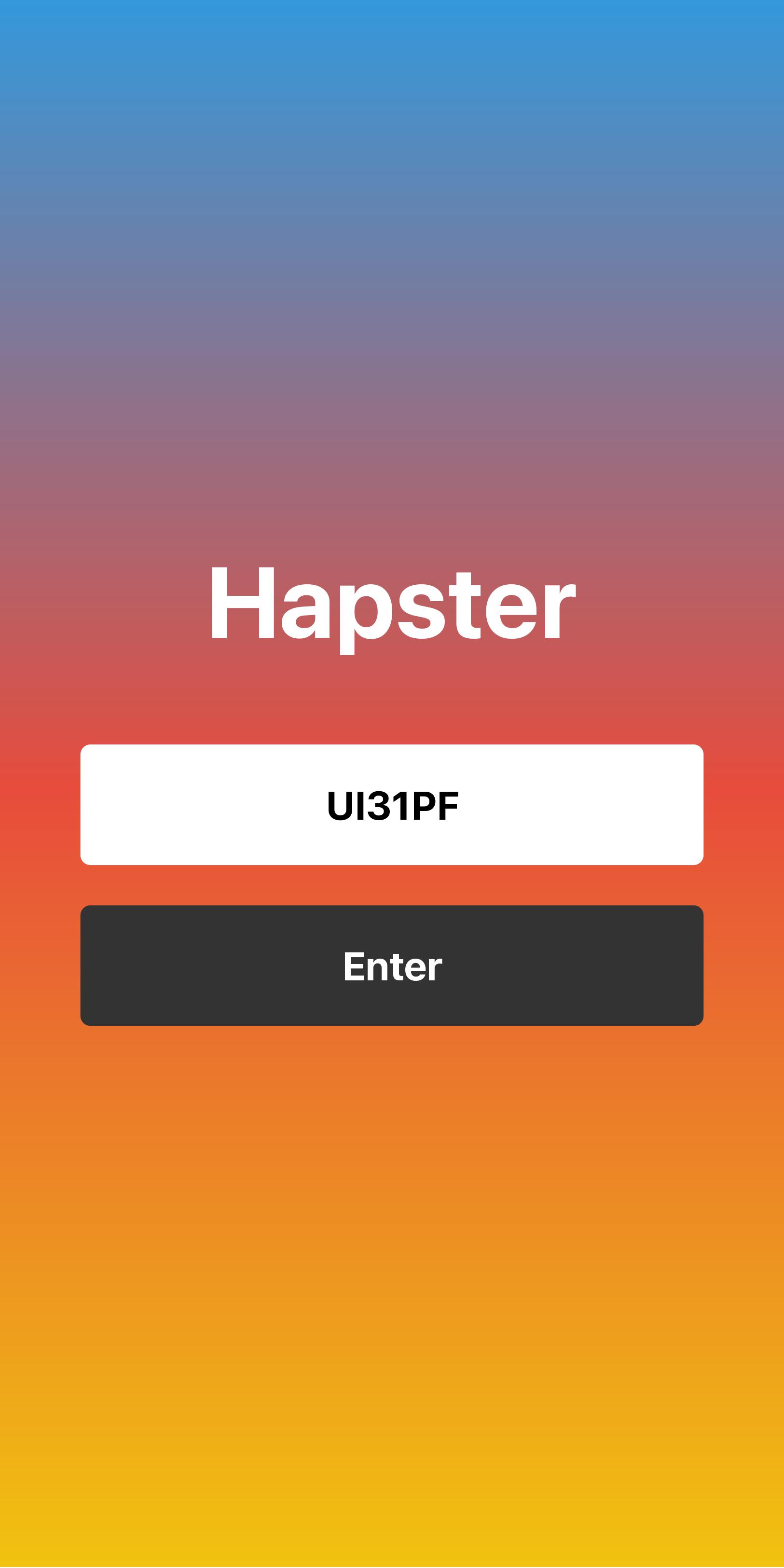}
        \Description[App main screen]{Hapster title page, with 'UI31PF' in an input box. An 'Enter' button at the bottom.}
        \subcaption{}
        \label{webapp1}
    \end{minipage}%
    \begin{minipage}{0.22\textwidth}
        \centering
        \includegraphics[scale=0.25]{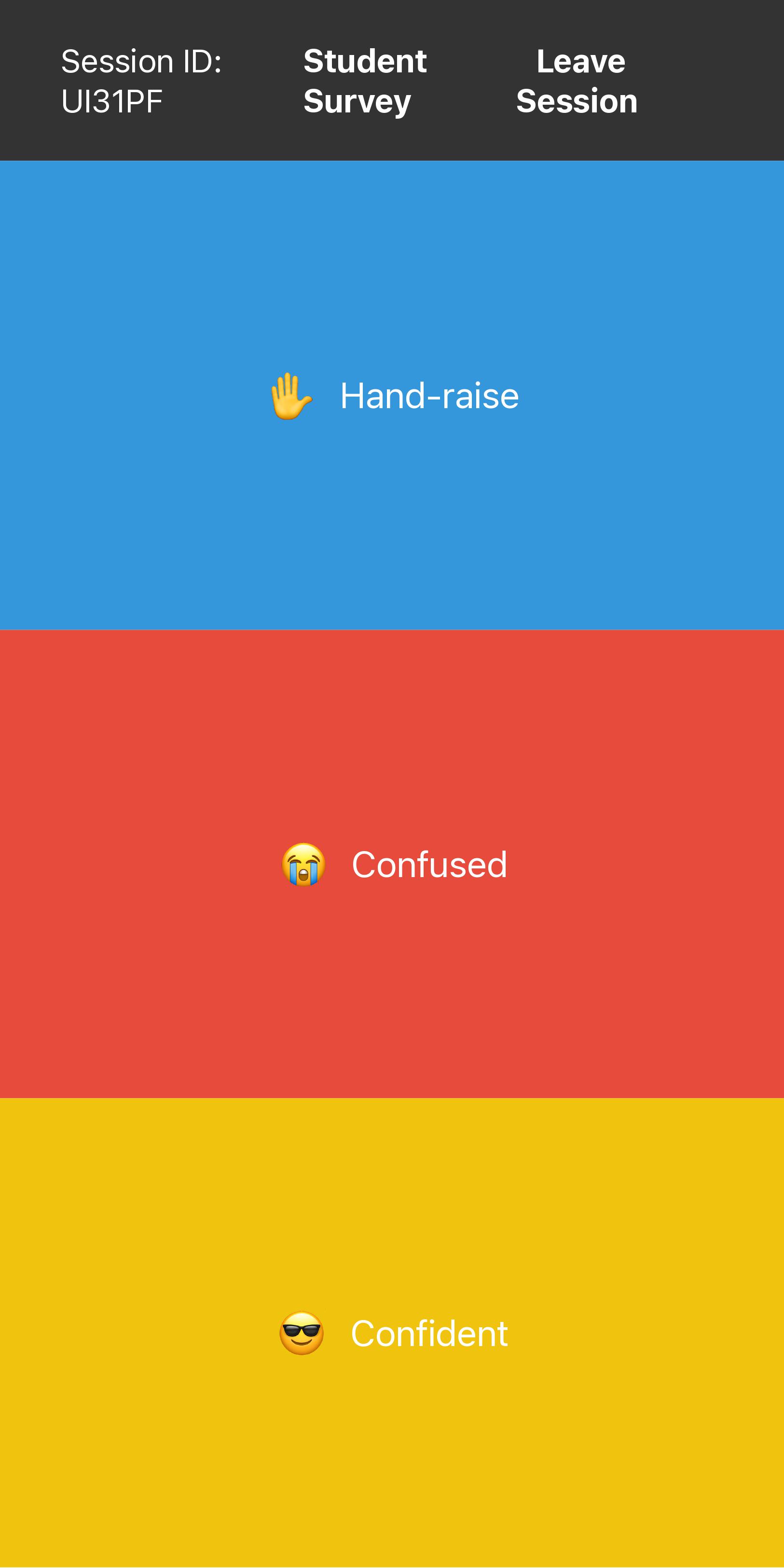}
        \Description[Reaction buttons screen]{Three buttons aligned vertically on screen: blue with Hand-Raise text and hand emoji, red with Confused text and crying emoji, and yellow with Confident text and sunglasses emoji}
        \subcaption{}
        \label{webapp2}
    \end{minipage}%
    \begin{minipage}{0.22\textwidth}
        \centering
        \includegraphics[scale=0.25]{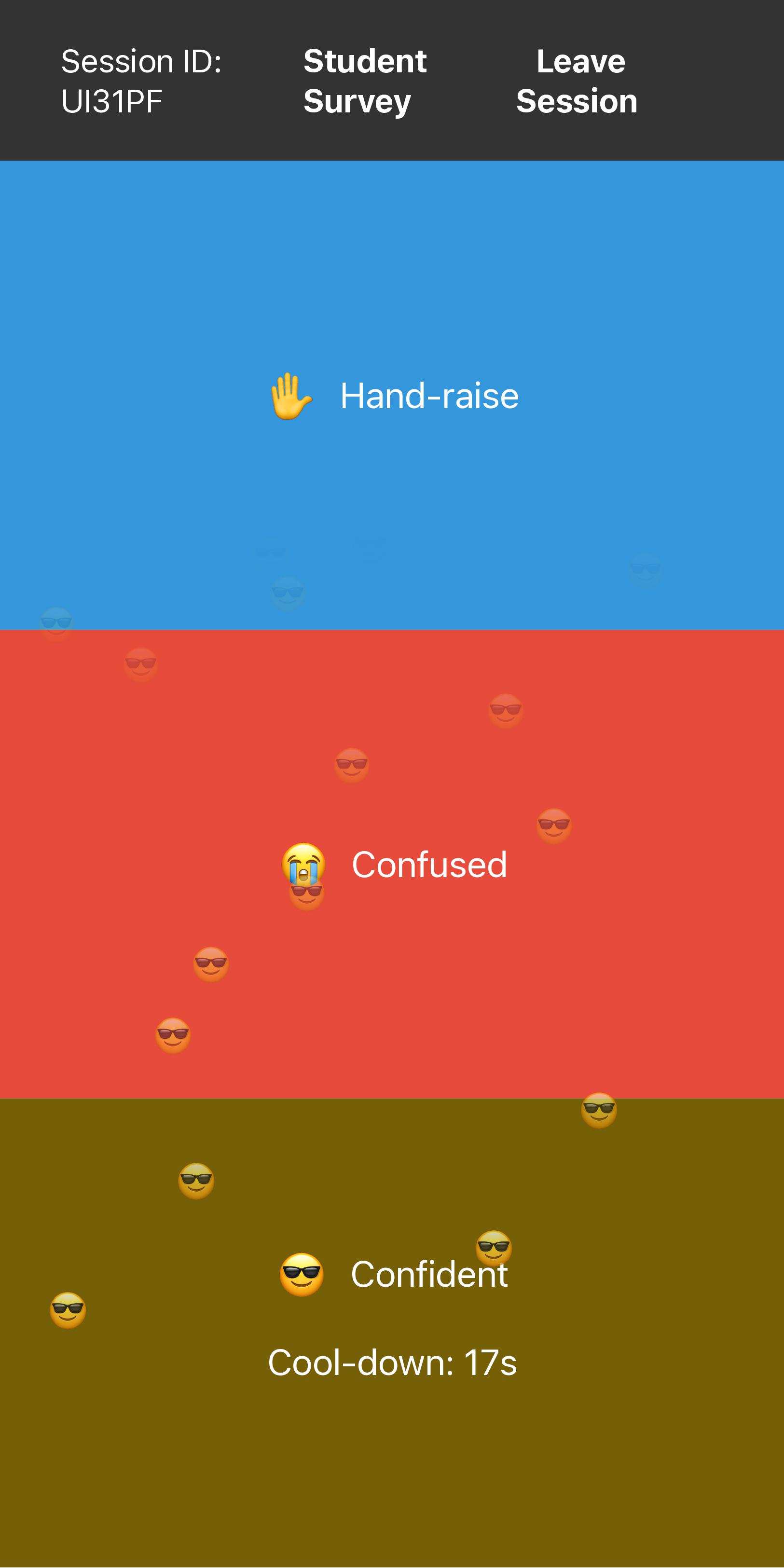}
        \Description[Pressed button screen]{Confident button is pressed, darkened slightly in color, with a countdown. Sunglasses emojis populate and fly in from the bottom and fade out at the top of screen.}
        \subcaption{}
        \label{webapp3}
    \end{minipage}%
    \begin{minipage}{0.22\textwidth}
        \centering
        \includegraphics[scale=0.25]{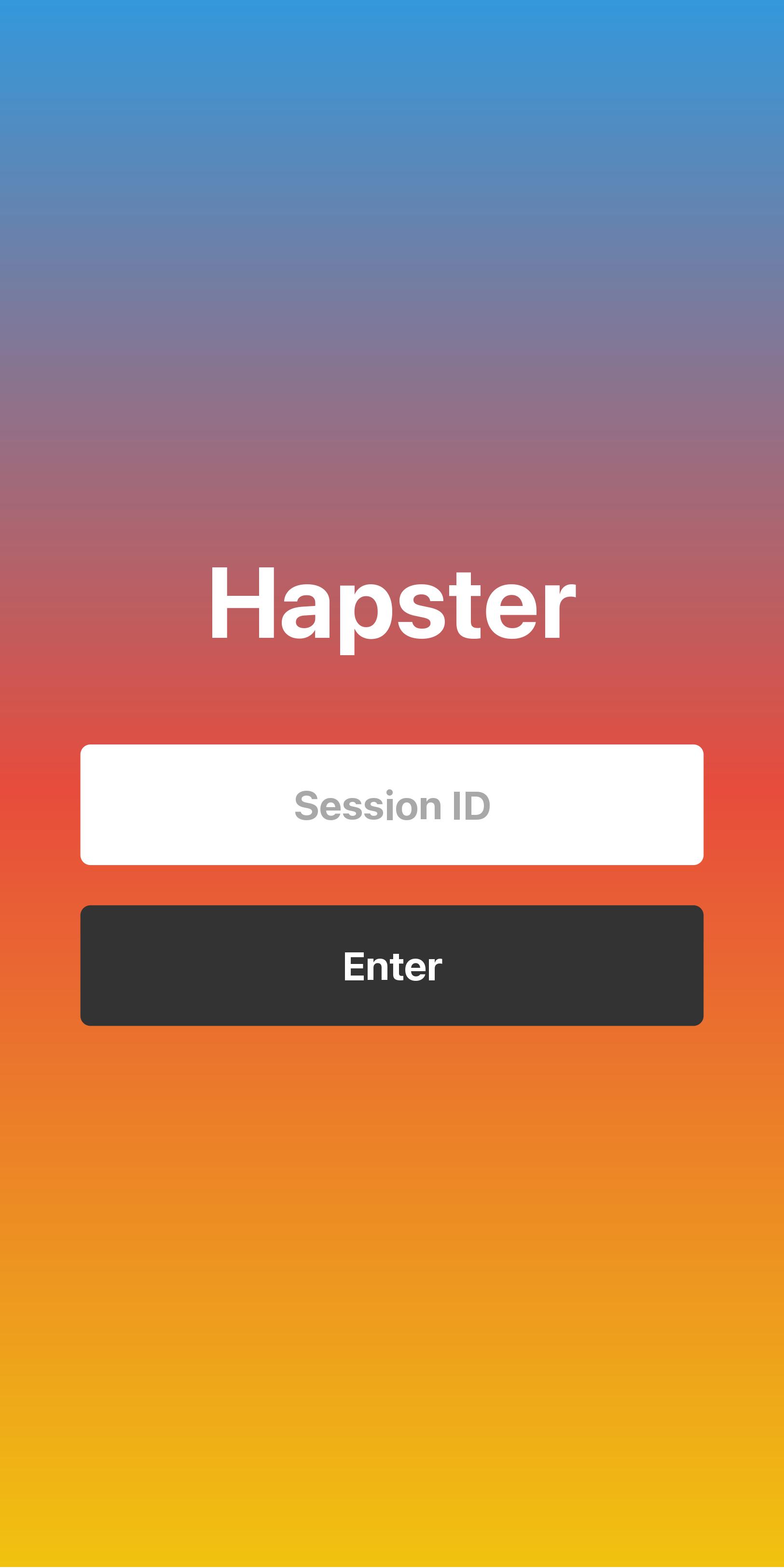}
        \Description[Return to app home screen]{Hapster title page, with an empty input box to enter a Session ID. An 'Enter' button at the bottom.}
        \subcaption{}
        \label{webapp4}
    \end{minipage}
\Description[Hapster student-facing web interface screens]{Hapster student-facing web interface screens as displayed on an iPhone device}
    \caption{Student-facing web interface. (a) After agreeing to the consent form, student enters a valid session ID on the home page to enter the main interface. (b) Student sees three reaction buttons: \textit{Hand-Raise}, \textit{Confused}, and \textit{Confident}. (c) Student taps the \textit{Confident} reaction, sending a stream of the associated emoji. (d) Student taps \textit{Leave Session} to return to the home page.}
    \label{webapp}
\end{figure*}

\subsection{Researcher-Facing Analytics Dashboard}
In order to conduct and monitor system trials, we built a dashboard (Figure \ref{researcher-dashboard}) featuring live reaction statistics and timeline to provide a detailed analysis of student engagement with the application during an active session. We used this interface during all system trials to detect when a new student reaction is sent and observe the instructor’s response to the reactions.

\setcounter{figure}{3}
\begin{figure}[H]
    \centering
    \includegraphics[scale=0.125]{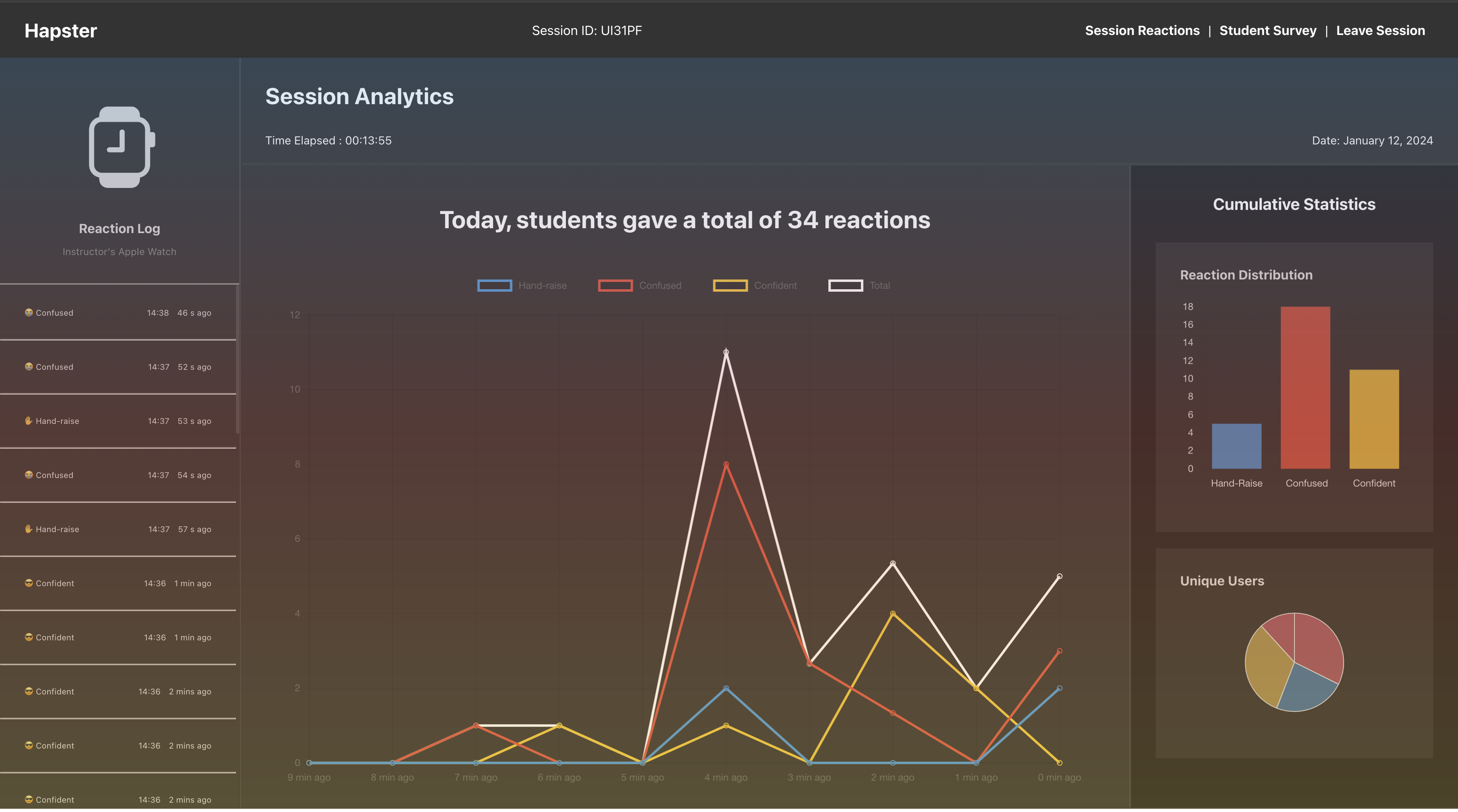}
    \Description[Hapster researcher interface screen]{Hapster researcher interface screen as displayed on a laptop device. Line graph in the middle peaks at 11 total reactions.}
    \caption{Researcher-facing web interface.}
    \label{researcher-dashboard}
\end{figure}

The \textit{Reaction Log} display presents a reverse-chronological sequence of student reactions. Each log entry includes the reaction icon, the corresponding label (\textit{Hand-raise}, \textit{Confused}, or \textit{Confident}), and the timestamp of the reaction. The line graph displays the trend of reactions aggregated by type over the last ten minutes. The bar chart offers a clear view of the cumulative distribution of various reactions since the start of the session. Finally, the pie chart breaks down reaction counts per unique anonymous user, helping evaluate individual user engagement. 

\section{Methods}
We evaluated the system by deploying it in 6 lectures at a U.S. university and soliciting data and observations from both instructors and students. All subjects were informed of the study and consented to our procedures before participating.

\subsection{Selecting Student Reactions}
\label{subsec:rxns}

We determined the three reactions by surveying 26 university instructors (9F, 17M) who taught a large lecture ($>60$ students) across academic disciplines during the fall 2023 semester. We asked the instructors to rank 12 given types of student reactions (described and justified in Appendix A.1) that they would find the most useful. Using the modified Borda count method, we found that the three most favored reactions were \textit{confusion} (37 points), \textit{hand-raising for questions} (29 points), and \textit{confidence in understanding} (21 points). Thus, we integrated these three reactions into Hapster.

\subsection{Pilot Study}
We ran a pilot study in a computer science lecture of 60 students to collect baseline data and test our system for potential issues. While there were no technical issues, the instructor noted that the haptic sequences could have been more intrusive and longer in duration in order to be more noticeable while the instructor was talking and moving around. Thus, we extended the number of times the haptic patterns repeated in each sequence to ensure that the instructor had enough time to recognize and respond to the sequences. The instructor also commented that the system helped them feel more connected with the students, as the instructor could more tangibly observe student engagement.

\subsection{In-Class System Studies}
We invited 6 of the instructors (2F, 4M) who participated in our initial survey to use our system during one of their regularly scheduled lectures, each lasting 50 or 80 minutes. As shown in Table \ref{recruitedclasses}, the recruited instructors covered a range of instructional methods, subjects, enrollment sizes, and student participation levels.

\begin{table*}[h]
\begin{tabular}{ l|l|l|l|l|l }
 Class & Course Type & Length & \thead[l]{Students \\ Present} & \thead[l]{Student \\ Participation Rate}  & Instruction Method \\ 
 \hline \hline
 
 C1 & First-year physical science & 80 min & 43 & 40\% & Writing on blackboard\\  
 C2 & Introductory social science & 50 min & 200 & 21\% & Writing on projected iPad \\
 C3 & Introductory quantitative math & 80 min & 60 & 28\% & Writing on blackboard\\
 C4 & \makecell[l]{Introductory quantitative \\ social science} & 80 min& 90 & 33\% & Projected slides \\
 C5 & First-year physical science & 80 min & 50 & 21\% & Writing on projected iPad \\
 C6 & Intermediate life science & 50 min & 70 & 46\% & Projected slides
\end{tabular}
\caption{Summary of recruited classes for system studies.}
\label{recruitedclasses}
\end{table*}

Student participation in the study was optional, with no penalty to their grade in the course, and the instructors did not know which students used the system. At least two members of our research team were present in the lecture room and wrote down observations for the duration of each study. The latter was not disclosed to the participants to eliminate bias. All data collected outside of the system-mediated data was recorded manually, and no video or audio was recorded at any point. Before each lecture, we provided the instructor with the Apple Watch with the Hapster application installed and trained the instructor to recognize the three haptic sequences by playing each sequence twice and asking the instructor to describe the sequences, encouraging them to consciously think about what differentiates them. We then assessed their identification accuracy and recall by asking them to identify nine haptic sequences (3 sequences x 3 repetitions in random order).

\subsection{Interviews and Surveys}

After each session, we interviewed the instructor to learn more about their perception of the system. We performed a thematic analysis of the interview transcripts. We also sent a survey to all students in the participating classes where we asked students to (1) comment on their lecture experience with the system, (2) compare our system with other SRSs they have used in the past, and (3) describe how they believe the system influenced their instructor's behaviour, with most questions either being open-ended or on a 5-point Likert scale. Additionally, we sent the NASA TLX questionnaire to both students and instructors to assess the workload of our system, with each question rated on a scale from 1 (extremely low) to 10 (extremely high). While instructors were not compensated, all students who interacted with the system during class and completed the post-study survey were entered into a raffle (if they opted in) for one \$10 Amazon gift card per class. 

\section{Results and Discussion}

\subsection{Usage patterns}
Though we generally received consistent feedback and commentary about our system from both the instructors and the students, we observed some unexpected behaviors that illuminated the limitations of our system.

\subsubsection{Instructor Haptics Test}
Most instructors (5 out of 6) were able to quickly memorize and later recognize the three distinct haptic sequences and their associated reactions, with the identification accuracy rates shown in Figure \ref{confusionmat}. All instructors, except P4, identified the \textit{Hand-Raise} sequence with 100\% accuracy. 4 of the 6 instructors commented that the sequences for \textit{Hand-Raise} and \textit{Confused} were similar and thus ambiguous. Though all instructors could verbally distinguish the \textit{Confident} sequence with ease, instructors misinterpreted \textit{Confused} as \textit{Hand-Raise} and \textit{Confident} equally, suggesting that more distinct haptic sequences or more training is necessary to further improve retention and differentiation rates. 

Despite receiving the same training as all other instructors, instructor P4 was unable to distinguish or identify any of the 3 sequences, and during the lecture, the instructor did not engage with the watch in any way despite students continuously sending reactions. The disconnect between the instructor and the students was reflected in the students' responses, all agreeing that their perceived effectiveness of Hapster was inconclusive due to the instructor’s lack of engagement with the system. In our debrief, the instructor acknowledged that they ``couldn't sense the difference in vibration sequences" and lamented that the haptics were “distracting” such that if they looked at the watch, they would lose their train of thought. Thus, from this session, we recognize that the effectiveness of haptics is dependent at least in part on the individual’s sensitivity to sensory stimuli.

\setcounter{figure}{5}
\begin{figure}[!htb]
    \centering
    \includegraphics[scale=0.4]{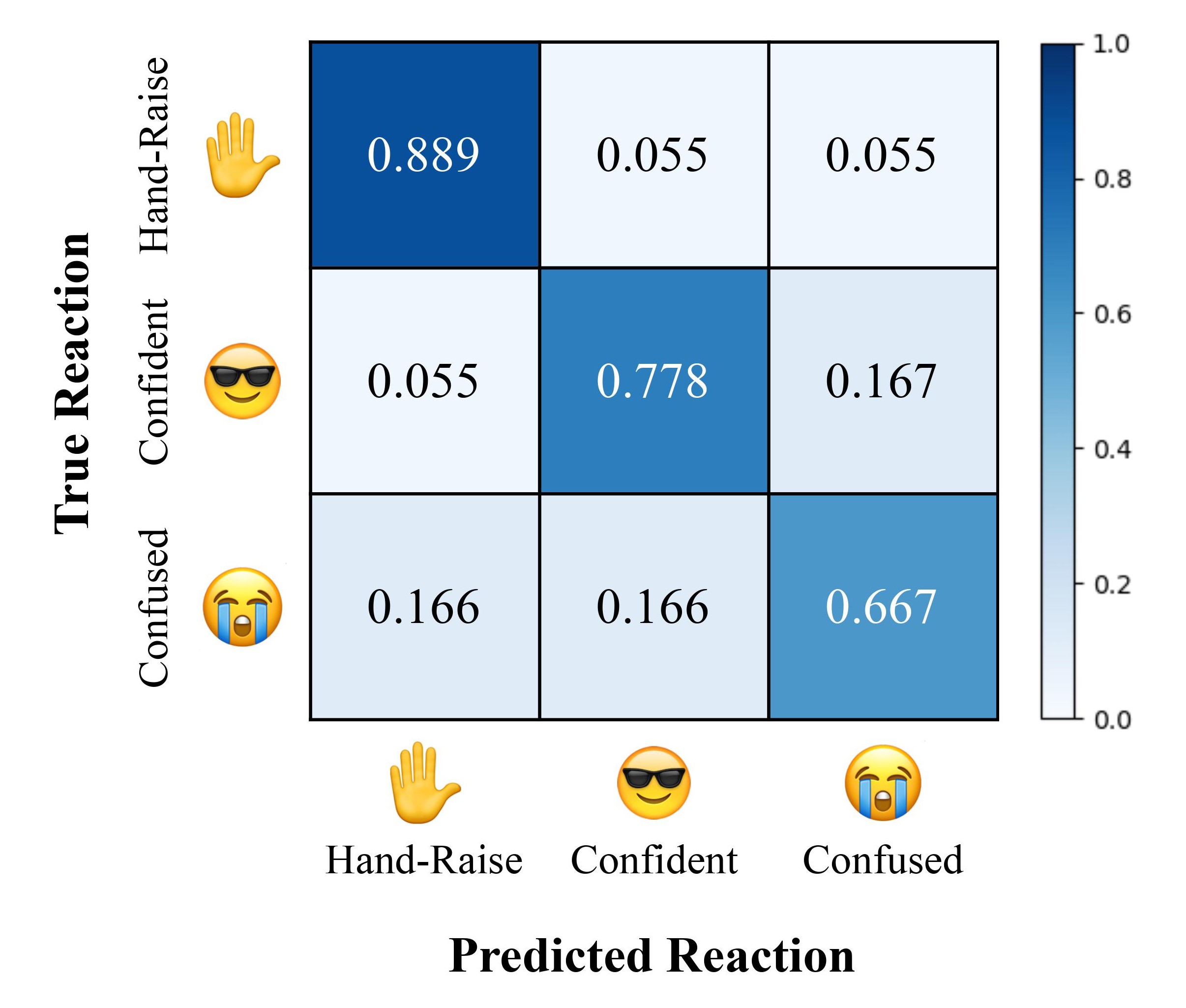}
    \Description[3-by-3 confusion matrix]{A 3-by-3 confusion matrix with the Confused, Confident, and Hand-Raise text and emojis as the cell labels, with \textit{True reaction} on the y-axis and \textit{Predicted Reaction} on the x-axis. Cells are shaded on a blue gradient according to value between 0 and 1. In order, from left to right, top to bottom, the values of the cells are 0.889, 0.055, 0.055, 0.055, 0.778, 0.167, 0.166, 0.166, and 0.667. The diagonal is shaded a darker blue according to the values.}
    \caption{Normalized confusion matrix for the instructor haptics test.}
    \label{confusionmat}
\end{figure}

\subsubsection{Student Abuse of System}
In class C2, we observed spamming behavior where a small subset of students continuously sent reactions and thus used the system disproportionately. In particular, a single student accounted for 30\% of all reactions in the class of 200 students, with 42 active participants using our system. Consequently, we implemented a cooldown feature, requiring users to wait 20 seconds before submitting another reaction of the same type. However, even with this preventative measure, classes C3 and C5 still had a few students who consistently submitted reactions every 20 seconds. In C3, the instructor would pause to ask for questions after receiving multiple \textit{Hand-Raise} and \textit{Confused} reactions, yet no students would ask questions or express any confusion. Instructor P5 was puzzled when P5 received \textit{Confused} reactions despite not having ``really said anything of substance.''  As these behaviors likely do not align with natural engagement with class content, we attribute these reactions to spammers rather than engaged students who want to continuously give feedback.


\subsection{Instructor Usage Feedback}
\textit{\textbf{Direct awareness and bi-directional communication.}} All instructors noted that the haptics increased their sense of connection with the students. 5 instructors reported increased awareness of the students' reactions, moving from traditional one-way communication to a bi-directional exchange. One instructor noted that the system \textit{``made [them] more aware [of the students' engagement levels] and increased the connection between [them] and the students, which is usually a one-way thing''} [P6]. Another instructor shared a similar experience: \textit{``I was discussing a concept they should already know, and the confidence reactions I received were mildly reassuring''} [P5].

\textit{\textbf{Live adjustment.}} 5 out of the 6 instructors commented that receiving indicators of the students' confidence and confusion levels allowed them to adjust their lecture pacing with minimal delay. One instructor shared how they slowed the lecture pace when they received confused reactions and sped up when they received confident reactions. Another instructor echoed this sentiment: \textit{``Knowing [the students are] confident means, I can move on and go faster''} [P3].

\textit{\textbf{First-time instruction.}} Two instructors hypothesized that the system would be most effective for an instructor teaching a course for the first time, so that timely feedback could best help improve the teaching curriculum: \textit{``if you're teaching a new class where you don’t have the experience, I think the tool would be awesome… knowing what's confusing and what is not confusing is super valuable''} [P1]. 

\textit{\textbf{Visual versus haptic notifications.}} Instructors had mixed appreciation for the visual notification that accompanied the haptic sequences. Overall, 4 out of the 6 instructors found the visuals to be helpful, especially when they missed the haptic sequence or failed to recognize the reaction in the moment. With that being said, one instructor strongly preferred haptics: \textit{``the visual is hard, and I don't think I looked at the watch at all''} [P5].

\subsection{Student Usage Feedback}

From a total of 72 student responses to our survey, we removed 7 responses that did not consent to using our system and 3 responses that claimed to have used our system in a course that we did not test in. Thus, we analyzed 62 valid student responses. Most questions were asked on a 5-point Likert scale from 1 (strongly disagree) to 5 (strongly agree). 

Consistent with prior research \cite{CallOnMe}, students who said they do not usually engage with their instructors during lectures cited not wanting to be a distraction, being shy, and general discomfort with asking questions in the moment. However, after using Hapster, about half of student responses cited an improvement in their learning experience since they were able to communicate their reactions anonymously. As one student put it, \textit{"I was able to convey my confidence, confusion, and whether or not I had a question without having to actively say something out loud"} [S57].

Students agreed that Hapster allowed them to convey more information during lectures to their instructor than they otherwise would have without the system (M = 3.58, SD = 1.02) and made them more willing to interact with their instructors in the moment (M = 3.71, SD = 1.03). Students also felt comfortable using the system during class (M = 3.65, SD = 1.06), but they were generally neutral about feeling more engaged with the lecture because of the system (M = 3.27, SD = 0.93). Students agreed that the system interface was visually appealing (M = 3.61, SD = 1.01). 

\subsection{Workload Assessment}
Students found our system less cognitively taxing than the instructors did. On a 10-point Likert scale, students rated the system low on mental (M = 2.42, SD = 2.02) and physical demand (M = 1.87, SD = 1.70). While instructors agreed that the system was not physically demanding (M = 1.50, SD = 0.84), there was more disparity in the mental demand (M = 4.33, SD = 3.20), likely due to the cognitive effort of lecturing while trying to recognize the haptic sequences.
 
\section{Limitations and Future Work}
As the effectiveness of our system relies on the instructors' ability to perceive and differentiate the haptic sequences, more research is needed to determine the types of sequences that would be most easily discernible, and if the benefits of multi-actuator devices \cite{xu2024haptics} could translate well to a classroom environment. Through a more flexible and programmable platform than the Apple Watch, the potential of Hapster can be further explored by building and using an in-house multi-actuator wrist-worn device. Parameters to investigate include actuator activation patterns, frequency, amplitude, rhythm, waveform, and duration.

The system itself would benefit from multiple improvements. First, while we iteratively improved the system to be more robust to high traffic and student abuse, spamming was still an issue at some points in the later sessions. Possible solutions include (1) limiting each unique user to a certain number of reactions per session or (2) banning overly-active users after a warning when they exceed a rate of 4 reactions per every 5 minutes (threshold based on our observations). Second, using a physical clicker-type device with no visual interface could be easier for students to use and introduce less distractions to student learning as opposed to using a personal electronic device. Third, the system could allow instructors to pick their own reactions and associated haptic sequences, allowing for more customization of the system to specific user preferences. Fourth, we could vary the intensity of a haptic sequence proportional to the aggregated count of the associated reaction.

Finally, more exploration of multi-modal interactions through integrating other types of wearables could improve the accessibility of our system for users with varying degrees of physical, mental, and sensory capabilities.

\section{Conclusion}
Instructors in in-person classrooms face challenges in receiving live, anonymized student feedback. Consequently, we identified an opportunity to build Hapster, a novel student response system with integrated haptic feedback, allowing instructors to receive live, anonymized, and aggregated student feedback without the need to divert their attention to a traditional visual interface. We conducted six in-class studies to evaluate our system as part of a multi-stage, survey- and classroom-based study. We found that Hapster was overall effective at facilitating student communication with instructors, increasing student engagement, and enhancing the instructors' understanding of their classes. Thus, this system positively impacted the students' learning experience by enabling instructors to quickly adapt to student needs. 

\bibliographystyle{ACM-Reference-Format}
\bibliography{references}

\appendix 

\section{RESEARCH METHODS}

\subsection{Instructor Pre-Survey: Reaction Preferences}

In the initial instructor survey, we provided the following 12 student engagement indicators for the instructors to rank their top three:

\begin{itemize}
    \item Hand-raising for questions/comments: in case people are not in your field of view, if you are facing away from the class while writing on the blackboard, etc.
    \item Speed up lecture pace: spend less time talking about this topic, move on to the next topic
    \item Slow down lecture pace: explain more in detail, spend more time on this topic
    \item Confused / frustrated: topic is still confusing
    \item Bored / distracted: lecture is boring, unable to focus
    \item Engaged: following what you are teaching
    \item In agreement: agree with the presented sentiments
    \item In disagreement: disagree with the presented sentiments
    \item Confident in understanding: this topic is clear, can move on to the next topic
    \item Curious / surprised: unexpected point raised, would like to think about it more
    \item Eager to participate: student has a thought to share
    \item Overwhelmed / need a break: tired, unable to focus, need a 5-minute break or stretch
\end{itemize}

\noindent Using the modified Borda count method, the results for the instructors' votes were as follows (we implemented the top three): 

\begin{enumerate}
    \item Confused / frustrated: 37 points
    \item Hand-raising for questions/comments: 29 points
    \item Confident in understanding: 21 points
    \item Engaged: 19 points 
    \item Curious / surprised: 14 points
    \item Speed up lecture pace: 11 points
    \item Eager to participate: 8 points
    \item Slow down lecture pace: 6 points 
    \item Bored / distracted: 5 points
    \item Overwhelmed / need a break: 3 points
    \item In disagreement: 2 points
    \item In agreement: 1 point
\end{enumerate}

\end{document}